\documentclass[conference]{IEEEtran}
\usepackage{cite}
\usepackage{amsmath,amssymb,amsfonts}
\usepackage{graphicx}
\usepackage{booktabs}
\usepackage{balance}
\usepackage[hidelinks]{hyperref}
\usepackage{eso-pic}
\AddToShipoutPictureBG*{%
  \AtPageLowerLeft{%
    \raisebox{18pt}{\makebox[\paperwidth][c]{\parbox{0.8\paperwidth}{\centering\scriptsize
    \copyright~2026 IEEE. Personal use of this material is permitted. Permission from IEEE must be obtained for all other uses, in any current or future media, including reprinting/republishing this material for advertising or promotional purposes, creating new collective works, for resale or redistribution to servers or lists, or reuse of any copyrighted component of this work in other works.}}}}}

\newcommand{\coes}{CoES}
\newcommand{\rh}[1]{\par\smallskip\noindent\textbf{#1}\enspace\ignorespaces}

\begin{document}

\title{Co-Evolution Signatures: Relational Timing Predicts User Activation}

\author{\IEEEauthorblockN{Naser Al-Obeidat}
\IEEEauthorblockA{\textit{Dept. Electrical and Computer Engineering}\\
\textit{Ingenuity Labs, Queen's University}\\
Kingston, ON, Canada\\
21nao3@queensu.ca}
\and
\IEEEauthorblockN{Michael Greenspan}
\IEEEauthorblockA{\textit{Dept. Electrical and Computer Engineering}\\
\textit{Ingenuity Labs, Queen's University}\\
Kingston, ON, Canada\\
michael.greenspan@queensu.ca}
}

\maketitle

\begin{abstract}
Predicting whether a user will be active on an online platform usually draws on two sources of information: the user's own recent activity and the current activity of the people they interact with. These sources describe the users on either side of a relationship, but not the relationship itself. In particular, two users can have a consistent temporal pattern: they may tend to become active together, one may reliably become active before the other, or activity by one may predict that the other will become more active shortly afterward. These patterns belong to how the two users behave relative to each other over time and therefore cannot be recovered from statistics computed from either user's history alone. We introduce Co-Evolution Signatures (CoES), a 24-dimensional, training-free representation that measures this temporal structure directly from interaction timestamps. Every quantity is computed from interaction metadata alone, using only the identities of the two endpoints and the timestamp of each event, with no message content, user profiles, or pretrained embeddings, and the prediction model, feature preprocessing, and train--test split are held identical across all feature sets, so measured gains are attributable to the representation rather than to any change in the learning setup. For each relationship, CoES measures synchrony, lead--follow ordering, activation lift, and response dynamics, and then summarizes these measurements across a user's strongest relationships. Because interactions shared by two users appear in both of their activity records and can create artificial coupling, we remove these shared events before computing CoES, separating how a relationship behaves over time from how frequently the pair interacts. We evaluate CoES on five public temporal interaction networks spanning private messaging, institutional email, and online question-and-answer communities. Adding CoES to a strong model that already observes a user's recent activity and their contacts' current activity improves next-window activation prediction on four of the five networks. On the two largest networks, CoES adds more predictive information than the contacts' current activity, improving AUC by $+0.018$ versus $+0.008$ on SuperUser and by $+0.014$ versus $+0.006$ on AskUbuntu. Finally, shifting one user's activity series in time while preserving that user's individual activity statistics destroys the relational signal, showing that the predictive information lies in how the two users' activity evolves together rather than in either user alone.
\end{abstract}

\begin{IEEEkeywords}
temporal networks, user modeling, temporal alignment, relational dynamics, activity prediction, social networks
\end{IEEEkeywords}

\section{Introduction}
Suppose we want to predict whether a user will be active tomorrow, and consider the two natural places one might look. The first is the user's own history, which records how often they act, how bursty their behavior is, whether their activity is rising or falling, and how long it has been since they last appeared. The second is the people they interact with, and specifically how many of them are active right now and how strongly.

Notice what both approaches do with the user's relationships, which is to treat them as plumbing. A relationship is the reason that someone else's activity is worth observing, but the relationship itself is never described, measured, or represented. This paper argues that relationships deserve to be objects of representation in their own right, because each one has a temporal character that carries predictive information which neither the user's history nor their contacts' current activity can supply.

By the temporal character of a relationship, we mean how the two sides of it move relative to each other in time, so that in some relationships one user reliably becomes active shortly before the other, in some the two move together, and in others one consistently trails. This character is genuinely a property of the relationship rather than of either participant. Consider two users who are identical on every individual statistic, having the same activity volume, the same contact count, the same burstiness, the same trend, and the same autocorrelation. If the first user's relationships place them ahead of the people they interact with while the second user's relationships place them behind, every individual statistic describes the two users identically, yet they occupy opposite temporal roles in their networks, and the difference is stored nowhere in either user's own history because it exists only in their relationships.

Snapshot features of the neighborhood cannot recover it either, as can be seen by supposing that a user's contacts are highly active at prediction time. For a user whose relationships historically run contact-first this strongly predicts activation, whereas for a user whose relationships run user-first it predicts much less, even though the snapshot is identical in both cases. What differs is the character of the relationships through which that snapshot should be interpreted.

The broader literature already indicates that relationships are temporally structured objects, since human communication arrives in bursts~\cite{r1,r2,r3,r4}, the timing of ties shapes how information spreads~\cite{r5,r6,r7,r8}, and behavior evolves together with network structure~\cite{r9,r10,r11}. Compact predictive representations of users, however, are still built one user at a time~\cite{r12,r13}, so the temporal character of their relationships has nowhere to live.

We introduce \coes{}, a fixed-length representation of that character, and evaluate it on next-window activation, which asks whether user $i$ will be active in window $t+1$ given everything observed up to window $t$. Our contributions are \coes{} itself, computed from timestamps alone with no training; an exclusive-series construction that separates a relationship's character from its intensity; evidence that relationships add signal beyond both self-history and current contact activity, and that on the two largest networks they carry more than the contact snapshot; and time-shift controls that localize the signal, together with an inexpensive statistic predicting in advance how much a network's relationships are worth representing.

\section{Related Work}
Temporal-network analysis describes a single user through inter-event burstiness~\cite{r1,r2,r3,r4}, rhythms and trends~\cite{r14,r15,r16,r17}, recency, and autocorrelation, all of which are standard in engagement and churn prediction~\cite{r12,r13,r18}, and all of which live inside one trajectory, so they form the individual-level part of our comparison. A separate literature establishes that relationships are dynamic objects, evolving through shared contexts and triadic closure~\cite{r9,r19,r20}, forming in ways that are themselves predictable~\cite{r21}, reinforcing behavioral similarity~\cite{r10,r22}, persisting with a characteristic strength and turnover~\cite{r23,r24,r25,r26}, and shaping propagation through their timing~\cite{r5,r6,r7,r8}. Graph representation learning encodes relational information into embeddings, capturing structural position~\cite{r27,r28}, interaction ordering~\cite{r29,r30}, and evolving node states from timestamped edge streams~\cite{r31,r32,r33,r34,r35}, surveyed in~\cite{r36} and benchmarked in~\cite{r37,r38}, but absorbs that information into an opaque learned space, whereas we isolate one named property without representation training, message content, or a learned embedding. Point-process models capture mutual excitation between event streams~\cite{r39,r40,r41} and, in joint formulations, let diffusion and network structure co-evolve~\cite{r42,r43}, but they fit parametric intensities rather than emit a fixed-length descriptor of a tie. Temporal dependence within relationships is also central to work on social influence~\cite{r44,r45,r46}, to actor-based models separating selection from influence~\cite{r11,r47}, and to Granger-style tests of directional predictability~\cite{r48,r49,r50}, though our aim is representational rather than causal, in that we record that one side systematically precedes the other while remaining agnostic about the mechanism.

\section{Co-Evolution Signatures}

\subsection*{Setup and Exclusive Series}
We observe timestamped interactions $\mathcal{E}=\{(u_m, v_m, t_m)\}$, in which each element records that user $u_m$ interacted with user $v_m$ at time $t_m$, over a period divided into $T$ equal windows. For each user $i$, let $a_i(t)$ be the number of interactions involving $i$ in window $t$, counted across all of the people they interact with. Because these counts are heavy-tailed~\cite{r1,r3}, we use the log-stabilized series $x_i(t)=\log(1+a_i(t))$, in which the $1+$ keeps empty windows well defined. Each user is represented through their partner set $P_i$, containing their top $K=20$ partners by interaction count and retaining only relationships with at least $\theta=3$ shared interactions, so that occasional contacts do not enter.

A relationship has two properties that must be kept apart, namely its \emph{intensity}, which is how often the pair interact directly~\cite{r23,r26}, and its \emph{character}, which is how their behavior aligns in time. Raw data conflates them, because if $i$ and $j$ exchange a message in window $t$ then that single event increments both $a_i(t)$ and $a_j(t)$, so intense relationships produce correlated series by construction and any coupling measured on raw series is partly a restatement of intensity. We therefore remove the shared events before measuring anything. Letting $m_{ij}(t)$ be the number of events in window $t$ involving both $i$ and $j$, the \emph{exclusive series} is
\begin{equation}
x_i^{-j}(t) = \log(1 + a_i(t) - m_{ij}(t))\,,
\label{eq:excl}
\end{equation}
in which the superscript $-j$ reads ``excluding $j$'', and $x_j^{-i}(t)$ exchanges the roles of $i$ and $j$ while subtracting the same quantity $m_{ij}(t)$. Since events with $j$ are a subset of events involving $i$, we have $m_{ij}(t)\le a_i(t)$ and the series is always well defined. The interpretation is direct, in that $x_i^{-j}$ is user $i$'s activity with everyone except $j$, so any coupling surviving this removal describes how the two users' broader behavior moves together with the direct channel between them taken out.

\subsection*{Four Measures of a Relationship's Character}
For a relationship $(i,j)$ we abbreviate the two exclusive series as $x\equiv x_i^{-j}$ and $y\equiv x_j^{-i}$, each of which is a sequence of $T$ numbers, and we form binary indicators $b_x=\mathbf{1}[x>0]$ and $b_y=\mathbf{1}[y>0]$, in which $\mathbf{1}[\cdot]$ returns $1$ when its condition holds and $0$ otherwise, so that $b_x$ records whether $i$ was active in a window rather than how much. Throughout, $\rho(\cdot,\cdot)$ denotes Pearson correlation, which lies in $[-1,1]$, and $x(a{:}b)$ denotes the entries of $x$ from position $a$ to position $b$.

\rh{Synchrony.} The first measure is $s_{ij}=\rho(x,y)$, set to $0$ if either series is constant since correlation is undefined against a constant, and a high value describes a relationship whose two sides are busy in the same windows and quiet in the same windows. Because $\rho(x,y)=\rho(y,x)$, synchrony is symmetric and therefore cannot indicate which side moves first, which is the role of the next measure.

\rh{Lead and follow.} We correlate the two series at offsets $\ell\in\{1,2,4\}$ windows in both directions:
\begin{align}
d_{ij}(\ell) &= \rho(x(1{:}T-\ell),\; y(1+\ell{:}T))\,, \label{eq:lead}\\
f_{ij}(\ell) &= \rho(y(1{:}T-\ell),\; x(1+\ell{:}T))\,. \label{eq:follow}
\end{align}
Equation~\eqref{eq:lead} pairs each earlier value of $x$ with the value of $y$ exactly $\ell$ windows later, so that a high $d_{ij}(\ell)$ describes a relationship in which changes in $i$ are followed $\ell$ windows later by changes in $j$, while equation~\eqref{eq:follow} reverses the roles and a high $f_{ij}(\ell)$ describes the opposite ordering. Using three offsets keeps the description multiscale without committing to a single delay~\cite{r17}.

\rh{Lift.} In sparse networks, whether a user appears at all can matter more than how much they do, so this measure uses the binary indicators. Co-activation lift compares observed joint presence against what independence would predict, and trigger lift makes the same comparison directional and forward-looking:
\begin{align}
L_{ij} &= \log\frac{\Pr[b_x b_y = 1]+\epsilon}{\Pr[b_x=1]\Pr[b_y=1]+\epsilon}, \label{eq:lift}\\
\tau_{i\to j} &= \log\frac{\Pr[b_y(t+1)=1 \mid b_x(t)=1]+\epsilon}{\Pr[b_y=1]+\epsilon}. \label{eq:trig}
\end{align}
Here $\Pr[\cdot]$ is a probability estimated by counting windows, the product $b_x b_y$ equals $1$ only when both sides are active, the bar $\mid$ denotes conditioning, and $\epsilon>0$ is a small constant preventing $\log 0$. The denominator of \eqref{eq:lift} is the joint rate expected if the two sides were unrelated, so $L_{ij}>0$ means the pair co-occur more often than chance and $L_{ij}<0$ less often, while the numerator of \eqref{eq:trig} is the fraction of windows following an active window of $i$ in which $j$ is active, measured against $j$'s overall activity rate. The reverse direction $\tau_{j\to i}$ is computed analogously.

\rh{Response gain.} The measures above compare activity levels, whereas this one relates a level on one side to a change on the other. Writing $\Delta y(t)=y(t)-y(t-1)$ for the window-over-window change in $y$,
\begin{equation}
g_{i\to j} = \rho(x(1{:}T-1),\; \Delta y(2{:}T))\,,
\label{eq:gain}
\end{equation}
which pairs $i$'s level in each window with $j$'s change into the next, so that a positive value means busy periods for $i$ tend to come just before $j$ ramps up while a negative value means they come before $j$ winds down. Lead and follow cannot separate these two cases, since both of them place $i$ ahead of $j$.

\subsection*{Pooling, the Pacemaker Score, and Cost}
Users differ in how many partners they have~\cite{r25}, so we pool the relationship-level statistics above into a fixed-length vector. Taken over the relationships in $P_i$, these are the mean, maximum, and standard deviation of synchrony (3); the mean and maximum co-activation lift (2); the mean trigger lift in each direction plus their difference (3); the mean response gain in each direction plus their difference (3); the log-count of strongly coupled relationships, meaning those with $|s_{ij}|>0.2$ (1); and, for each offset $\ell\in\{1,2,4\}$, the mean lead, maximum lead, mean follow, and mean lead minus mean follow (12). This yields 24 numbers, in which the means and maxima describe how strongly the user's relationships are coupled while the directional differences describe which side of those relationships the user occupies.

One coordinate deserves particular attention. Writing $\overline{d_i(\ell)}$ and $\overline{f_i(\ell)}$ for the lead and follow values averaged over user $i$'s partners, the one-window asymmetry $p_i=\overline{d_i(1)}-\overline{f_i(1)}$ is what we call the \emph{pacemaker score}, which is positive for users sitting on the leading side of their relationships and negative for those on the trailing side. Subtracting cancels the overall coupling magnitude and leaves only the asymmetry, which is a quantity that cannot be defined for a user in isolation.

Computing one signature requires examining at most $K$ relationships over $T$ windows, so for $N$ eligible users the full build costs $O(|\mathcal{E}|+NKT)$ operations after binning the stream, where $|\mathcal{E}|$ is the number of events and both terms are linear in the data. MathOverflow, with roughly 507k interactions and 24.8k users, builds in under a minute on a single CPU core. Because everything reduces to windowed counts and moments, the same statistics can be maintained incrementally over a rolling history, so the method applies to streaming data without training or embeddings, in contrast to models that must be refit as the graph evolves~\cite{r31,r32,r35}.

\section{Experimental Setup}
\rh{Datasets and features.} We use five public SNAP tempo\-ral networks~\cite{r51,r52,r53}. CollegeMsg contains 59,835 private messages exchanged on a UC-Irvine platform over 194 days, binned into daily windows, and email-Eu-core contains 329,910 e-mails within a European research institution over a contiguous 526-day active stretch, divided into 260 windows of roughly two days. MathOverflow, SuperUser, and AskUbuntu contain answers, comments, and comment replies from three Stack Exchange sites, with 506,550, 1,443,339, and 964,423 interactions over 6.4, 7.6, and 6.2 years, binned into 340, 360, and 320 approximately weekly windows, and for AskUbuntu we remove a 156-day pre-launch interval containing only 14 events. Each dataset's window size is fixed once and used throughout the study~\cite{r14}. Alongside the 24 \coes{} features we use four \textsc{Stat} features, namely log activity volume, log partner count, out-interaction ratio, and active-window fraction, together with six \textsc{Self} features, namely inter-event burstiness $(\sigma-\mu)/(\sigma+\mu)$~\cite{r2,r4}, in which $\mu$ and $\sigma$ are the mean and standard deviation of the gaps between a user's consecutive events, along with trend, recency, and autocorrelations at offsets 1, 2, and 4.

\rh{Task and baseline.} The target is whether user $i$ is active in the next window, written $y_{i,t}=\mathbf{1}[a_i(t+1)>0]$. The \textsc{Hist} group supplies eight causal measurements of the user at time $t$, meaning measurements that use no information after $t$, and these are the activity in windows $t$ through $t-3$, the trailing four- and eight-window activity sums, the trailing active-window fraction, and the time since last activity. The \textsc{PState} group supplies eight measurements of the neighborhood at the same moment, namely the fraction of top partners active at $t$, the total, maximum, and tie-weighted partner activity at $t$, the total partner activity at $t-1$, and the count, total activity, and maximum activity of partners who interacted with $i$ in the preceding eight windows. The baseline therefore already knows what the user has been doing and what the people they interact with are doing now, so that \coes{} supplies only the character of the relationships connecting the two.

\rh{Leakage, models, and uncertainty.} Partner sets and \coes{} signatures are estimated entirely from the first half of each dataset, models are trained on the 50--80\% portion of the timeline, and evaluation uses the final 20\%, so every signature is frozen before the prediction period begins and no coordinate uses a future window, an ordering that recent benchmarking work identifies as essential for temporal tasks~\cite{r37,r38}. We evaluate with logistic regression on standardized features~\cite{r54}, using $C=1$ for the inverse regularization strength, and with histogram gradient-boosted trees using 300 iterations, early stopping, and $\ell_2=1$, with settings held identical across all feature combinations so that differences reflect the features rather than separate tuning. Because each user contributes many examples which are not independent of one another, we bootstrap 500 times while resampling whole users rather than individual examples~\cite{r55}, and report ROC--AUC~\cite{r56} changes with 95\% user-clustered intervals.

\begin{figure}[t]
\centering
\includegraphics[width=\columnwidth]{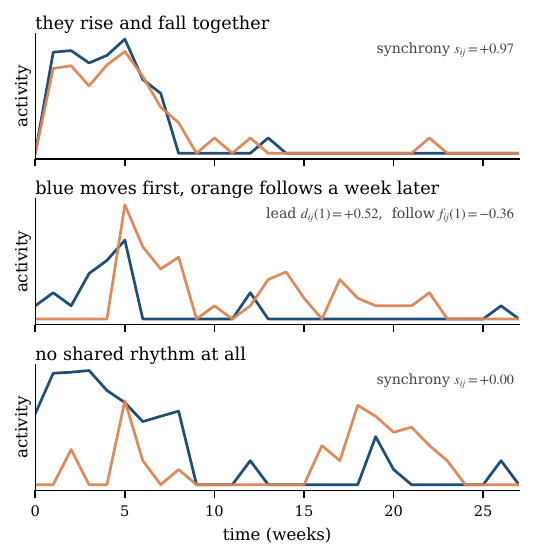}
\caption{Three real relationships from CollegeMsg, each drawn as its two exclusive weekly activity series. The top pair moves together, in the middle pair one side consistently moves a week before the other, and the bottom pair shares no rhythm. These are the three temporal characters that \coes{} measures.}
\label{fig:rel}
\end{figure}

\section{Results}
\rh{What a relationship's character looks like.} Fig.~\ref{fig:rel} shows the raw material of the representation on three relationships from CollegeMsg. In the first, the two users rise and fall in the same weeks, with synchrony $s_{ij}=0.97$. In the second, one user's activity consistently precedes the other's by about a week, with $d_{ij}(1)=+0.52$ against $f_{ij}(1)=-0.36$, an ordering that synchrony alone cannot record. In the third, the two series share no rhythm and every coupling measure is near zero. \coes{} assigns each relationship its position among these characters and pools them per user.

\rh{Relationships add signal beyond self-history and contact state.} The central question is whether relationship character adds anything once the model already knows both the user's recent activity and the neighborhood's current state, so the baseline to beat is \textsc{Hist+PState}, and it does. Adding \coes{} improves activation prediction by $\Delta\mathrm{AUC}=+0.007$ on email-Eu with a confidence interval of $[+0.005,+0.008]$, by $+0.004$ on MathOverflow with $[+0.003,+0.004]$, by $+0.008$ on SuperUser with $[+0.007,+0.009]$, and by $+0.007$ on AskUbuntu with $[+0.006,+0.008]$, so that every interval excludes zero. On CollegeMsg, which is the smallest and shortest network, the gain is $+0.001$ with an interval spanning zero and we report no effect. Since the baseline already observes whether the user was active, how recently, whether the people they interact with are active, and how much, the gain on the remaining four networks must come from something that neither side of the relationship supplies on its own.

\rh{On large networks, relationship history beats the contact snapshot.} Starting from \textsc{Hist} alone, \textsc{PState} adds $+0.008$ AUC on SuperUser while \coes{} adds $+0.018$, and on AskUbuntu \textsc{PState} adds $+0.006$ while \coes{} adds $+0.014$, so on both of the largest networks the history of a user's relationships is more informative than the current state of the people they interact with. The reason is the earlier point seen from the other side, in that two users can sit in equally busy neighborhoods while their relationships run in opposite directions, so \textsc{PState} describes both environments identically while \coes{} tells the model how each user's relationships translate that environment into activation. On AskUbuntu the effect is sharper still, since \textsc{Hist+CoES} matches the full \textsc{Hist+PState+CoES} model, meaning that the relationships already contain the activation signal carried by generic contact-state measurements and more besides. We note that \coes{} contributes 24 features against \textsc{PState}'s eight, so part of this margin may reflect capacity rather than information.

\rh{Character is not intensity in disguise.} The exclusive-series step matters most in finely binned networks, where a single shared event visibly moves both users' counts. On email-Eu, synchrony computed from raw series correlates with tie weight~\cite{r23} at $r=0.45$ and on CollegeMsg at $r=0.26$, but after removing the pair's shared events both correlations fall to $r=0.16$, showing that a large share of the raw coupling in these networks was simply intensity. In the weekly networks the conflation is mild, because one mutual event is a tiny fraction of a week's activity, and inclusive and exclusive synchrony agree to within $0.01$. Predictive performance is preserved under the exclusive construction, so the step removes the intensity component without removing the useful one, and this gives the representation a precise meaning in which a \coes{} relation describes how $i$'s activity with everyone else moves relative to $j$'s activity with everyone else.

\begin{table*}[t]
\caption{Activation forecasting: decomposing partner information (HGB AUC, temporal train/test split). \textsc{PState} contains generic partner-state variables (partner activity now and recently); \textsc{CoES} is the 24-dimensional co-evolution signature estimated on the first half of the span. $\Delta$ is the increment of adding \textsc{CoES} on top of \textsc{Self+PState}, with a 95\% CI from a user-clustered bootstrap (500 resamples); $^{*}$ marks CIs excluding zero.}
\label{tab:ta}
\centering
\small
\begin{tabular}{lcccccc}
\toprule
Dataset & Base & \textsc{Self} & \textsc{+PState} & \textsc{+CoES} & \textsc{+PState+CoES} & $\Delta$\textsc{CoES}$\mid$\textsc{PState} \\
\midrule
CollegeMsg & 0.03 & 0.899 & 0.898 & 0.899 & \textbf{0.899} & +0.001\,{\scriptsize[-0.003,+0.004]} \\
email-Eu & 0.56 & 0.793 & 0.806 & 0.801 & \textbf{0.813} & +0.007\,{\scriptsize[+0.005,+0.008]}$^{*}$ \\
MathOverflow & 0.07 & 0.922 & 0.924 & \textbf{0.928} & 0.928 & +0.004\,{\scriptsize[+0.003,+0.004]}$^{*}$ \\
SuperUser & 0.05 & 0.829 & 0.837 & \textbf{0.847} & 0.846 & +0.008\,{\scriptsize[+0.007,+0.009]}$^{*}$ \\
AskUbuntu & 0.04 & 0.865 & 0.871 & \textbf{0.879} & 0.878 & +0.007\,{\scriptsize[+0.006,+0.008]}$^{*}$ \\
\bottomrule
\end{tabular}
\end{table*}

\rh{Time-shift control and alignment strength.} For each true relationship we circularly shift one exclusive series in time, wrapping its end around to its start, which preserves the series' total activity, marginal distribution, sparsity, burst structure, and autocorrelation while destroying only the temporal correspondence between the two sides, in the spirit of randomized reference models for temporal networks~\cite{r6,r14}. After the shift, coupling collapses toward zero on every network, so that each user considered alone is statistically unchanged, only the relationship's alignment was severed, and the signal went with it, which means that what \coes{} measures is a property of the relationship rather than of either participant. The size of that collapse is itself informative, so we define \emph{alignment strength} as $A=\mathbb{E}[s_{ij}^{\mathrm{partner}}]-\mathbb{E}[s_{ij}^{\mathrm{shifted}}]$, in which $\mathbb{E}[\cdot]$ averages over sampled pairs and the two terms are synchrony measured on true relationships and on their time-shifted controls, so that the difference isolates genuine relational alignment from chance correlation between unrelated series. Across the five datasets $A$ tracks the \coes{} activation gain, though with only five points this is suggestive rather than conclusive. Because $A$ can be estimated cheaply from a sample of pairs before any further construction, it answers the deployment question in advance.

\begin{figure}[t]
\centering
\includegraphics[width=\columnwidth]{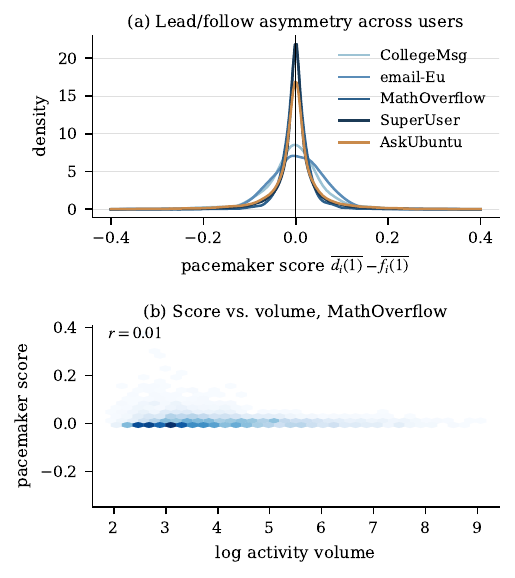}
\caption{(a) Distribution of the one-window lead and follow asymmetry, the pacemaker score, across eligible users. (b) Pacemaker score against log activity volume on MathOverflow, with $r=0.01$.}
\label{fig:pace}
\end{figure}

\rh{The pacemaker score is a relational role.} Fig.~\ref{fig:pace}(a) shows that $p_i$ is concentrated near zero but carries substantial tails in both directions in every dataset, so leading and trailing are both common relational roles rather than quirks of a few unusual accounts. The score is also nearly independent of activity scale, since its absolute correlation with activity volume stays below $0.09$ across all five datasets and is $r=0.01$ on MathOverflow as shown in Fig.~\ref{fig:pace}(b), meaning that heavy users are no more likely to lead their relationships than light ones. We test this independence more strongly by attempting to predict $p_i$ from the full \textsc{Stat+Self} representation, which comprises ten individual features, and with both ridge regression and gradient-boosted trees the cross-validated $R^2$ is at or below zero on all five datasets, meaning that the fitted models do no better on held-out users than always predicting the mean. A user's individual statistics therefore carry essentially no information about the role they play in their relationships.

\section{Discussion and Conclusion}
The experiments support a three-way decomposition of temporal user behavior into the user's own state, the neighborhood's current state, and the character of the relationships connecting them. These sources are not substitutes, because a relationship determines what the other two mean, in that a busy neighborhood predicts activation for a user whose relationships run contact-first and predicts far less for one whose relationships run user-first, while \textsc{PState} reports the same condition in both cases, and this is why \coes{} continues to help when \textsc{Hist} and \textsc{PState} are both present. The time-shift control localizes the signal, since shifting one series preserves its values, distribution, total, and internal temporal structure while severing only the alignment, yet the signal disappears anyway, which places the information in how the two trajectories are arranged relative to each other in time. Every coordinate remains readable in these terms, and because the pacemaker score is nearly uncorrelated with ordinary activity statistics, this relational role constitutes a genuinely additional behavioral dimension rather than a rephrasing of volume or regularity.

The method requires only timestamped interactions and simple operations on binned counts, with no message content, profiles, pretrained embeddings, neural training, or specialized graph infrastructure, and alignment strength serves as a cheap screening test beforehand. \coes{} captures the predictive character of relationships and does not claim causation, since an ordering may reflect influence, shared schedules, common interests, or external events~\cite{r16,r22,r44,r45,r46}, and disentangling these remains difficult even with purpose-built models~\cite{r11,r47}, so we measure that it exists and predicts rather than why it arises. We chose linear correlations, binary lift, and equal-width windows for their cost and readability, and the framework extends naturally to nonlinear dependence measures~\cite{r57,r58}, adaptive time scales~\cite{r16,r17}, higher-order neighborhood relations~\cite{r7,r59,r60}, and learned relational operators~\cite{r39,r40}, each of which changes how a relationship's character is measured while preserving the object of interest. A direct comparison against learned temporal graph embeddings~\cite{r31,r32,r33,r36}, which we position as complementary rather than competing, is left to future work.\\
Individual history describes what the user has been doing, contact state describes what is happening around them, and co-evolution describes the temporal character of the relationships connecting the two, which is a piece that becomes visible only when relationships, rather than users, are taken as objects of representation.

\vspace{-1.35pt}
\bibliographystyle{IEEEtran}
\bibliography{refs}

\end{document}